\documentstyle[12pt,epsf]{article}
\def\beq{\begin{equation}}
\def\eeq{\end{equation}}
\def\beqn{\begin{eqnarray}}
\def\eeqn{\end{eqnarray}}

\def\ie{{\it i.e.,\ }}

\def\nicefrac#1#2{\hbox{${#1\over #2}$}}

\def\gev{{\rm\,GeV}}

\def\mev{{\rm\,MeV}}

\begin{document}

\begin{titlepage}

\setcounter{page}{1}

\rightline{JLAB--THY-98-06}
\rightline{UFIFT--HEP-97-25}

\vfill
\begin{center}
 {\Large \bf Drell-Yan Non-Singlet Spin Cross Sections \\ and Spin Asymmetry to $O(\alpha_s^2)$ (II)}

\vfill
\vfill
 {\large Sanghyeon Chang$^{*}$\footnote{
        E-mail address: schang@phys.ufl.edu},
        Claudio Corian\`{o}$^{**}$\footnote{
        E-mail address: coriano@jlabs2.jlab.org}   
        and R. D. Field$^{*}$\footnote{
        E-mail address: rfield@phys.ufl.edu}}
\\
\vspace{.12in}
 {\it $^{*}$   Institute for Fundamental Theory, Department of Physics, \\
        University of Florida, Gainesville, FL 32611, 
        USA}
\\
\vspace{.12in}
{\it $^{**}$ Jefferson Lab, Newport News, VA, 23606, USA}
\vspace{.075in}

\end{center}
\vfill
\begin{abstract}
We present predictions for the non-singlet 
Drell-Yan longitudinal spin cross sections and spin asymmetry, $A_{LL}$, 
in proton-proton collisions at large $p_T$ at the RHIC energy of $200\gev$ at 
next-to-leading order QCD.  The higher order corrections to the non-singlet 
polarized cross section, $\sigma_{NS}^{LL}$, are sizeable and similar to those 
found for the unpolarized cross section.
The non-singlet asymmetry parameter, $A^{NS}_{LL}$, is very stable against 
higher order corrections and is a direct measurement of the 
non-singlet (\ie valence) polarized quark distributions within the proton.   
\end{abstract}

\end{titlepage}

\setcounter{footnote}{0}

\newcommand{\beqa}{\begin{eqnarray}}
\newcommand{\eeqa}{\end{eqnarray}}
\newcommand{\eps}{\epsilon}

\section{Introduction}
The study of the spin content of the nucleon at hadron/hadron colliders is 
one of the most interesting aspects of QCD which is increasing receiving 
attention both from the theoretical and from the experimental side. 
 
Among the most interesting questions that a $p$-$p$ collider can help us answer 
is the issue of the distribution of the proton spin among its constituents and 
the measurement of various polarized parton distributions. In this new setting, 
the presence of a direct gluon coupling, absent in polarized 
Deep Inelastic scattering processes, will allow a direct test of the gluon 
contribution to the polarization of the 
nucleon, with measurements covering a wide $x$ (Bjorken-x) range. 
We mention that, especially at smaller $x$ values, the gluon polarization 
$\Delta G$ is
strongly model-dependent and, therefore, poorly known from DIS experiments alone. 

Along these lines, in the last few years, several studies regarding the analysis 
of the scaling violations in polarized $p$-$p$ collisions have 
been presented, and considerable progress has been made in the theoretical 
investigation of various important polarized processes. 
For instance, it has become now clear that inclusive prompt photon 
production \cite{GVetal} $(p\,\,p\to \gamma + X)$, photon-plus-jet production $(p\,\, p\to \gamma + J)$ \cite{CG}, 
and polarized production of 2 jets $(p\,\,p\to J\,\, J)$ are among the most important 
processes to be studied at RHIC, the Relativistic Heavy Ion Collider at Brookhaven. 

The corresponding cross sections show sizable asymmetries (see for instance \cite{CG1}) 
with good hope for the experimental program. We mention also that, beside the usual asymmetries 
for total cross sections or large $p_T$ cross sections, other observables of the final state 
can be used as a way to test the size and the sign of $\Delta G$. 

In this respect, the study of the event-structure of the final state and 
the asymmetries for various correlations 
(for instance rapidity correlations, angular correlations in the photon/jet system) 
have been shown also to be  a source of important information on the 
polarized parton distributions 
and can help us discriminate among the various different models \cite{CG1}.  

More recently, lepton pair production via the Drell-Yan mechanism has 
also received considerable attention. 
The process has been historically a clean way to test QCD and the parton 
model and continues to be so also 
in the case of polarized collisions. 
In a seminal paper \cite{RS}, Ralston and Soper pointed out that the factorization formula for parton distributions 
for Drell-Yan includes, beside the usual longitudinally polarized parton 
distributions, also parton 
distributions for transverse spin. 
In particular the study 
of the transverse spin distribution 
is one of the challenging tasks that the STAR and PHENIX collaborations at RHIC are going to undertake within the next few years. 

On the technical side, we remind that 
although formal proofs of factorization are still missing, direct calculations 
show that the Drell-Yan cross sections does factorize, as shown in nontrivial cases 
(for instance for the nonzero $p_T$ distributions \cite{CCE}). 
This opens the way to an accurate study of this important cross section both 
for longitudinal and transverse spin. At $p_T=0$, where $p_T$ describes the 
transverse momentum of the lepton pair, the calculations simplify drastically 
and various results have been reported in the literature 
(see for instance \cite{ratcliffe,TG} for the longitudinal case).

In previous works \cite{CCFG1}, 
we have analyzed the spin dependence of hard scattering in Drell-Yan (with 
longitudinal polarization) 
at parton level to NLO QCD, at nonzero $p_T$ of the lepton pair.  
In this work we proceed with the analysis at hadron level of 
these former results, thereby providing predictions for the behavior of the cross section at RHIC energies.  
Since our NLO analysis is limited to the non singlet sector, our study 
is exploratory in nature, and we have limited our attention to just one particular center of 
mass scattering energy ($200$ GeV), right in the middle of the planned RHIC energy range.  
We will also limit our interest to a single set of parton distributions, specified below.  
Our interest, in this work, is limited to assess the stability of the full 
$O(\alpha_s^2)$ calculation, 
predict the asymmetries of various helicity dependent amplitudes and show that the corresponding K factors are comparable to the unpolarized ones.  

The asymmetries turn out to be sizable and can be larger than what estimated here once the gluon contribution is fully included. We hope to return to a re-analysis of the complete process in the near future.

\section{Structure of the Differential Cross Section}

We begin by examining the contribution to the production of {\it virtual} photons with invariant mass, $Q$, in hadron-hadron collisions, 
\beq
H_1 + H_2\to \gamma^* + X,
\label{rdf_x1}
\eeq
from the
$2\to 3$ parton subprocess,
\beq
a+b\to \gamma^*+c+d.
\label{rdf_x2}
\eeq

Here $H_1$ and $H_2$ are incoming hadrons with 4-momenta, $P_1$ and $P_2$, respectively, and $q$ is the 4-momentum of the virtual photon, $\gamma^*$, as shown in Fig.~1.  The 4-momenta of the
incoming two parton $a$ and $b$ are labeled by $p_1$ and $p_2$ respectively, and the outgoing 4-momenta of partons $c$ and $d$ are labeled by $k_2$ and $k_3$.  The virtual photon 4-momenta is given by $k_1=q$.  Conservation of energy and momentum implies that
\beq
p_1+p_2=k_1+k_2+k_3.
\label{rdf_x3}
\eeq
The hadron-hadron process (\ref{rdf_x1}) is described in terms of the
invariants,
\beq
S=(P_1+P_2)^2,\qquad T=(P_1-q)^2,\qquad U=(P_2-q)^2.
\label{rdf_bigs}
\eeq
In addition, it is convenient to define the scaled variables
\beq
x_T=2q_T/W\ {\rm and}\ \tau=Q^2/S,
\eeq
where $W=\sqrt{S}$ is the hadron-hadron
center-of-mass energy and $q_T$ is the transverse momentum of the virtual photon with invariant mass, $Q$.  It is also useful to define the following two quantities,
\beq
\bar x_1={Q^2-U\over S}=\nicefrac12 e^y\sqrt{x_T^2+4\tau},\qquad
\bar x_2={Q^2-T\over S}=\nicefrac12 e^{-y}\sqrt{x_T^2+4\tau},
\label{rdf_xbar}
\eeq
where $y$ is the rapidity of the virtual photon.
The $2\to 3$ parton subprocess (\ref{rdf_x2}) is described in terms of the invariants,
\beq
s=(p_1+p_2)^2,\qquad t=(p_1-q)^2,\qquad u=(p_2-q)^2,
\label{rdf_stu}
\eeq
and
\beq
s_2=s_{23}=(k_2+k_3)^2,
\eeq
where momentum and energy conservation (\ref{rdf_x3}) implies
\beq
s+t+u=Q^2+s_2.
\label{rdf_sum1}
\eeq

In QCD the hadronic cross section is related to the parton subprocess according to
\begin{eqnarray}
\lefteqn{d\sigma(H_1+H_2\to \gamma^*+X;W,q_T,y)=} \nonumber \\
&&G_{H_1\to a}(x_1,M^2)dx_1 
G_{H_2\to b}(x_2,M^2)dx_2\left({d\hat\sigma\over dtdu}
(ab\to\gamma^*cd;s,t,u)\right) dtdu,
\label{rdf_eq1}
\end{eqnarray}
where $G_{H_1\to a}(x_1,M^2)dx_1$ is the number of partons of flavor $a$ with momentum fraction, 
$x_1=p_1/P_1$, within hadron $H_1$ at the factorization scale $M$.  Similarly, $G_{H_2\to b}(x_2,M^2)dx_2$ is the number of partons of flavor $b$ with momentum fraction $x_2=p_2/P_2$, 
within hadron $H_2$, at the factorization scale $M$.  In the remainder of this paper we will take the factorization scale to be $Q$ and evaluate the parton subprocesses at the same scale $Q$.  Using (\ref{rdf_bigs}) and (\ref{rdf_stu}) we see that
\begin{eqnarray}
s&=&x_1x_2S,\\
(t-Q^2)&=&x_1(T-Q^2)=-x_1\bar x_2S,\\
(u-Q^2)&=&x_2(U-Q^2)=-x_2\bar x_1S,
\end{eqnarray}
which from (\ref{rdf_sum1}) implies that
\beq
x_1x_2-x_1\bar x_2-x_2\bar x_1=\tau_2-\tau,
\label{rdf_sum2}
\eeq
where $\bar x_1$ and $\bar x_2$ are defined in (\ref{rdf_xbar})
and
\beq
\tau_2=s_2/S.
\eeq
It is now easy to compute the Jacobian
\beq
dx_2dt={s\over4(x_1-\bar x_1)}\ dx_T^2dy,
\eeq
which when inserted into (\ref{rdf_eq1}) and integrating over $x_1$ and
$s_2$ gives
\begin{eqnarray}
\lefteqn{S{d\sigma\over dq_T^2dy}(W,q_T,y)=
\int_{x_1^{min}}^1dx_1\int_0^{s_2^{max}}ds_2\ {1\over (x_1-\bar x_1)}} 
\nonumber \\
&&\ G_{H_1\to a}(x_1,Q^2) 
G_{H_2\to b}(x_2,Q^2)\ s{d\hat\sigma\over dtdu}
(ab\to\gamma^*cd;s,t,u),
\label{rdf_eq2}
\end{eqnarray}
where
\beq
x_2={x_1\bar x_2+\tau_2-\tau\over x_1-\bar x_1},
\eeq
and
\beq
s_2^{max}=A=(\tau-\bar x_1+x_1(1-\bar x_2))S,\qquad
x_1^{min}={\bar x_1-\tau\over 1-\bar x_2}.
\label{rdf_defa}
\eeq
The maximum value of $\tau_2$ arises when
$x_2=1$ in (\ref{rdf_sum2}), while the minimum value of $x_1$ occurs when $\tau_2=0$ and $x_2=1$. 

For $2\to 2$ parton subprocesses such as the Born contribution in Fig.~2 one has
\beq
s{d\hat\sigma\over dtdu}(s,t,u)=\delta(s_2)\ s{d\hat\sigma\over dt}
(a+b\to\gamma^*+c;s,t),
\eeq
which when inserted into (\ref{rdf_eq2}) results in
\begin{eqnarray}
\lefteqn{S{d\sigma\over dq_T^2dy}(W,q_T,y)=
\int_{x_1^{min}}^1dx_1\ {1\over (x_1-\bar x_1)}} 
\nonumber \\
&&\ G_{H_1\to a}(x_1,Q^2) 
G_{H_2\to b}(x_2,Q^2)\ s{d\hat\sigma\over dt}
(ab\to\gamma^*c;s,t),
\label{rdf_eq3}
\end{eqnarray}
where in this case $s_2=0$ and $s+t+u=Q^2$. 

The Drell-Yan differential cross section for producing a muon pair with transverse momentum $q_T$ and invariant mass, $Q$, at rapidity $y$ in the hadron-hadron collision,
\beq
H_1+H_2\to (\gamma^*\to \mu^+\mu^-)+X,
\eeq
at center-of-mass energy $W$, is given by
\begin{eqnarray}
\lefteqn{S{d\sigma\over dQ^2dq_T^2dy}(H_1H_2\to\mu^+\mu^-+X;W,Q^2,q_T,y)=} \nonumber \\
&&\quad\left({\alpha\over3\pi Q^2}\right)
S{d\sigma\over dq_T^2dy}(H_1H_2\to\gamma^*+X;W,q_T,y),
\label{rdf_new1}
\end{eqnarray}
where $\alpha$ is the electromagnetic fine structure constant and
where the virtual photon differential cross section is given by (\ref{rdf_eq2}) or (\ref{rdf_eq3}). 

\subsection{The Non-Singlet Cross Section}

We define the ``non-singlet" cross section at the hadron level to be the antihadron-hadron 
minus the hadron-hadron cross section as follows:
\beq
\sigma_{NS}=\sigma_{\overline H_1H_2}-\sigma_{H_1H_2},
\eeq
and at the parton level to be the quark-antiquark minus the quark-quark cross section,
\beq
{\hat\sigma}_{NS}={\hat\sigma}_{q\bar{q}}-{\hat\sigma}_{qq}.
\eeq
Since the antihadron-hadron cross section is larger, this definition yields positive values
for $\sigma_{NS}$.  For parton distributions, on the other hand, we use the customary non-singlet definition,
\beq
G_{H\to q}  ^{NS}=G_{H\to q} - G_{H \to \bar{q}}.
\eeq
In proton-proton collisions the non-singlet cross section is given by
\beqa
\lefteqn{\sigma^{NS}_{pp}=\sigma_{\bar{p}p}-\sigma_{pp}} \nonumber \\
&&\ =\left(G_{\bar{p}\to q}G_{p\to q} +G_{\bar{p}\to\bar{q}}G_{p\to \bar{q}}\right) \otimes\hat{\sigma}_{qq}  \nonumber \\
&&\ +\left(G_{\bar{p}\to q}G_{p\to \bar{q}}+G_{\bar{p}\to\bar{q}}G_{p\to q}\right) \otimes\hat{\sigma}_{q\bar{q}}  \nonumber \\
&&\ -\left(G_{p\to q} G_{p\to q}+G_{p\to\bar{q}}G_{p\to \bar{q}}\right) \otimes\hat{\sigma}_{qq}  \nonumber \\
&&\ -\left(G_{p\to q}G_{p\to \bar{q}}+G_{p\to\bar{q}}G_{p\to q}\right) \otimes\hat{\sigma}_{q\bar{q}}  \nonumber \\
&&\ =G_{p\to q}^{NS}G_{p\to q}^{NS}\otimes{\hat\sigma}_{qq}^{NS}, 
\eeqa
where the convolution $\otimes$ is defined in (\ref{rdf_eq2}) and (\ref{rdf_eq3}).

\subsection{The Spin Dependent Cross Sections}

The non-singlet proton-proton helicity dependent Drell-Yan cross section can be written in 
the form,
\begin{eqnarray}
\lefteqn{S{d\sigma_{NS}\over dQ^2dq_T^2dy}(W,Q^2,q_T,y,\Lambda_1,\Lambda_2)=} \nonumber \\
&&\left({\alpha\over3\pi Q^2}\right)\sum_q\sum_{\lambda_1,\lambda_2}
\int_{x_1^{min}}^1dx_1\int_0^{s_2^{max}}ds_2\ {1\over (x_1-\bar x_1)} \nonumber \\
&&\ G^{NS}_{p[\Lambda_1]\to q[\lambda_1]}(x_1,Q^2) 
G^{NS}_{p[\Lambda_2]\to q[\lambda_2]}(x_2,Q^2)\ s{d{\hat\sigma}_{NS}\over dtdu}(s,t,u,h),
\label{rdf_nsfh1}
\end{eqnarray}
where $\Lambda_1$ and $\Lambda_2$ are the helicities of the initial two hadrons, respectively, 
and $h=4\lambda_1\lambda_2$. 
In \ref{rdf_nsfh1} the sum is over the quark flavor and the quark helicities $\lambda_1$ and $\lambda_2$, while the non-singlet helicity structure functions are given by
\beq
G^{NS}_{p[\Lambda]\to q_f[\lambda]}=G_{p[\Lambda]\to q_f[\lambda]}
-G_{p[\Lambda]\to \bar q_f[\lambda]}.  
\eeq

For one quark flavor
\beqa
\sigma_{++}^{NS}=&\left(G_{\uparrow\to \uparrow}^{NS}G_{\uparrow\to\uparrow}^{NS}
+G_{\uparrow\to\downarrow}^{NS}
G_{\uparrow\to\downarrow}^{NS}\right)
\otimes {\hat\sigma}_{++}^{NS} \nonumber \\
&+ \left(G_{\uparrow\to\uparrow}^{NS}
G_{\uparrow\to\downarrow}^{NS}
+G_{\uparrow\to\downarrow}^{NS}
G_{\uparrow\to\uparrow}^{NS}\right)
\otimes {\hat\sigma}_{+-}^{NS},
\label{rdf_new2a}
\eeqa
and
\beqa
\sigma_{+-}^{NS}=&\left(G_{\uparrow\to\uparrow}^{NS}
G_{\downarrow\to\uparrow}^{NS}
+ G_{\uparrow\to\downarrow}^{NS}
G_{\downarrow\to\downarrow}^{NS}\right)
\otimes {\hat\sigma}_{++}^{NS} \nonumber \\
&+ \left(G_{\uparrow\to \uparrow}^{NS}
G_{\downarrow\to\downarrow}^{NS}
+G_{\uparrow\to \downarrow}^{NS}
G_{\downarrow\to\uparrow}^{NS}\right)
\otimes {\hat\sigma}_{+-}^{NS},
\label{rdf_new2b}
\eeqa
where $\uparrow$ and $\downarrow$ refer to helicity $+\nicefrac12$ and $-\nicefrac12$, 
respectively, and where we have used simplified ``convolution notation".
The unpolarized non-singlet cross section is the average of $\sigma_{++}^{NS}$ and
$\sigma_{+-}^{NS}$,
\beq
\sigma_{NS}^{\Sigma}=\nicefrac12\left(\sigma_{++}^{NS}+\sigma_{+-}^{NS}\right),
\eeq
while the non-singlet longitudinal spin difference (\ie polarized) 
cross section is defined by,
\beq
\sigma_{NS}^{LL}=\nicefrac12\left(\sigma_{++}^{NS}-\sigma_{+-}^{NS}\right).
\eeq
{}From (\ref{rdf_new2a}) and (\ref{rdf_new2b}) it is easy to show that, for one quark flavor,
\beq
\sigma_{NS}^{\Sigma}=\left(G_{p\to q}^{NS}G_{p\to q}^{NS}\right)\otimes {\hat\sigma}_{NS}^{\Sigma},
\eeq
and
\beq
\sigma_{NS}^{LL}=\left(\Delta G_{p\to q}^{NS}\Delta G_{p\to q}^{NS}\right)\otimes {\hat\sigma}_{NS}^{LL},
\eeq
where the unpolarized parton distributions is the sum of the two spin states,
\beq
G^{NS}_{p\to q}=G_{\uparrow\to \uparrow}^{NS}+G_{\uparrow\to \downarrow}^{NS},
\eeq
and the ``polarized" parton distributions is the difference of the two spin states,
\beq
\Delta G^{NS}_{p\to q}=G_{\uparrow\to \uparrow}^{NS}-G_{\uparrow\to \downarrow}^{NS}.
\eeq
Similarly, the parton unpolarized non-singlet cross section is the average of ${\hat\sigma}_{++}^{NS}$ and ${\hat\sigma}_{+-}^{NS}$,
\beq
{\hat\sigma}_{NS}^{\Sigma}=\nicefrac12\left({\hat\sigma}_{++}^{NS}+{\hat\sigma}_{+-}^{NS}\right),
\eeq
and the polarized parton non-singlet cross section is defined by,
\beq
{\hat\sigma}_{NS}^{LL}=\nicefrac12\left({\hat\sigma}_{++}^{NS}-{\hat\sigma}_{+-}^{NS}\right).
\eeq

In Ref.\cite{CCFG1} we showed that to order $\alpha_s^2$ the Drell-Yan non-singlet 
spin-dependent parton-parton differential in (\ref{rdf_nsfh1}) can be 
written as the Born term plus four ({\it unpolarized}) cross sections as follows:
\begin{eqnarray}
\lefteqn{s{d{\hat\sigma}_{NS}\over dtdu}(s,t,u,h)=
(1-h)\delta(s_2)s{d{\hat\sigma}^{\Sigma}_{B}\over dt} } \nonumber \\
&&+(1-h)s{d{\tilde\sigma}^{\Sigma}_{1}\over dtdu}+
(1-h)s{d{\hat\sigma}^{\Sigma}_{2}\over dtdu}+
(1-h)s{d{\hat\sigma}^{\Sigma}_{3}\over dtdu}-
(1+h)s{d{\hat\sigma}^{\Sigma}_{4}\over dtdu},
\label{rdf_nsfh2}
\end{eqnarray}
where $h=4\lambda_1\lambda_2$.  Hence, to order $\alpha_s^2$ the Drell-Yan non-singlet 
unpolarized differential cross section for producing muon pair with transverse 
momentum $q_T$ and invariant mass, $Q$, at rapidity $y$ in the proton-proton collision at 
center-of-mass energy, $W$, is given by,
\begin{eqnarray}
\lefteqn{S{d\sigma^{\Sigma}\over dQ^2dq_T^2dy}(W,Q^2,q_T,y)=} \nonumber \\
&&\left({\alpha\over3\pi Q^2}\right)\sum_{q=u,d}
\int_{x_1^{min}}^1dx_1\int_0^{s_2^{max}}ds_2\ 
{1\over (x_1-\bar x_1)} \nonumber \\
&&\quad G^{NS}_{p\to q}(x_1,Q^2) 
G^{NS}_{p\to q}(x_2,Q^2)\ s{d{\hat\sigma}^{\Sigma}_{NS}\over dtdu}(s,t,u),
\label{rdf_nsunpol}
\end{eqnarray}
where
\beq
s{d{\hat\sigma}^\Sigma_{NS}\over dtdu}(s,t,u)=
\delta(s_2)s{d{\hat\sigma}_{B}^\Sigma\over dt}+
s{d{\tilde\sigma}_{1}^\Sigma\over dtdu}+
s{d{\hat\sigma}_{2}^\Sigma\over dtdu}+
s{d{\hat\sigma}_{3}^\Sigma\over dtdu}-
s{d{\hat\sigma}_{4}^\Sigma\over dtdu},
\label{rdf_nspin2}
\eeq
and the polarized cross section is given by,
\begin{eqnarray}
\lefteqn{S{d\sigma^{LL}\over dQ^2dq_T^2dy}(W,Q^2,q_T,y)=} \nonumber \\
&&\left({\alpha\over3\pi Q^2}\right)\sum_{q=u,d}
\int_{x_1^{min}}^1dx_1\int_0^{s_2^{max}}ds_2\ {1\over (x_1-\bar x_1)} \nonumber \\
&&\quad\Delta G^{NS}_{p\to q}(x_1,Q^2) 
\Delta G^{NS}_{p\to q}(x_2,Q^2)\ s{d{\hat\sigma}^{LL}_{NS}\over dtdu}(s,t,u),
\label{rdf_nsddtll}
\end{eqnarray}
where
\beq
s{d{\hat\sigma}^{LL}_{NS}\over dtdu}(s,t,u)=
-\delta(s_2)s{d{\hat\sigma}_{B}^\Sigma\over dt}-
s{d{\tilde\sigma}_{1}^\Sigma\over dtdu}-
s{d{\hat\sigma}_{2}^\Sigma\over dtdu}-
s{d{\hat\sigma}_{3}^\Sigma\over dtdu}-
s{d{\hat\sigma}_{4}^\Sigma\over dtdu}.
\label{rdf_nspin3}
\eeq
The unpolarized Born term, ${\hat\sigma}^{\Sigma}_B$ and the four {\it spin averaged} 
cross sections,
${\tilde\sigma}_{1}^\Sigma$, ${\hat\sigma}_{2}^\Sigma$, ${\hat\sigma}_{4}^\Sigma$, 
and ${\hat\sigma}_{4}^\Sigma$, are given in Ref.\cite{CCFG1} and Ref.\cite{EMP}.

\section{Proton-Proton Collisions at 200 GeV}
\subsection{Polarized and Unpolarized Cross Sections}

To illustrate the effect of the higher order terms in muon-pair production
in hadron-hadron collisions, we compute the non-singlet polarized and unpolarized Drell-Yan
cross sections in proton-proton collisions at $W=200\gev$.  The non-singlet unpolarized cross 
section is given by (\ref{rdf_nsunpol}), which in convolution notation becomes
\beq
\sigma_{NS}^{\Sigma}=\sum_{q=u,d}\left(G_{p\to q}^{NS}G_{p\to q}^{NS}\right)\otimes {\hat\sigma}_{NS}^{\Sigma},
\label{rdf_conave1}
\eeq
where from (\ref{rdf_nspin2})
\beq
{\hat\sigma}^\Sigma_{NS}={\hat\sigma}_B+{\tilde\sigma}_1+{\hat\sigma}_2+{\hat\sigma}_3
-{\hat\sigma}_4.
\label{rdf_conave2}
\eeq
The non-singlet longitudinal ({\it polarized}) cross section given by (\ref{rdf_nsddtll}) becomes
\beq
\sigma_{NS}^{LL}=\sum_{q=u,d}\left(\Delta G_{p\to q}^{NS}\Delta G_{p\to q}^{NS}\right)\otimes {\hat\sigma}_{NS}^{LL},
\label{rdf_conll1}
\eeq
where from (\ref{rdf_nspin3})
\beq
{\hat\sigma}^{LL}_{NS}=-{\hat\sigma}_B-{\tilde\sigma}_1-{\hat\sigma}_2-{\hat\sigma}_3
-{\hat\sigma}_4.
\label{rdf_conll2}
\eeq
We use GS solution A \cite{GS} for the polarized and unpolarized parton 
distributions ($\Lambda=200\mev$) which
are shown in Fig.~3 at $Q=2\gev$.  Fig.~4 shows the polarized and unpolarized 
non-singlet cross sections to leading order ({\it Born term}) and to 
order $\alpha_s^2$ at $W=200\gev$, $Q=M=2\gev$, and $y=0$.
Fig.~4 shows the ``number density" $dN/d\tau dx_Tdy$ which is constructed from the 
cross section as follows:
\beq
{d\sigma\over d\tau dx_Tdy}={x_T\over2}\ S^2{d\sigma\over dQ^2dq_T^2dy},
\eeq
using an integrated luminosity of $100$/pb.  

Fig.~5 shows the non-singlet polarized and unpolarized 
cross sections to leading order and to order $\alpha_s^2$ at $W=200\gev$ and $Q=M=2\gev$
integrated over $y$
\beq
{d\sigma\over d\tau dx_T}=\int_{y_{min}}^{y_{max}}\ {d\sigma\over d\tau dx_Tdy}\ dy.
\eeq
Here the number density $dN/d\tau dx_T$ corresponds to the number of events per
unit $\tau$ per unit $x_T$ in $100$/pb.  To compute the number of events in a bin of
$\tau$ of size $0.0001$ (corresponding to $\Delta M=2\gev$ at $W=200\gev$) and a bin 
of $x_T$ of size $0.01$ (corresponding to $\Delta p_T=1\gev$ at $W=200\gev$), for example, 
one would multiply $dN/d\tau dx_T$ by $10^{-6}$.

Fig.~6 and Fig.~7 show the ratio of the full order $\alpha_s^2$ to the leading
order cross sections (\ie the ``K-factors") for the polarized and unpolarized case.  
As has been previously seen, the K-factor for the unpolarized cross section is quite large.
The polarized K-factor is almost the same as the unpolarized case.  The difference 
comes only from ${\hat\sigma}_4$ in (\ref{rdf_conave2}) and (\ref{rdf_conll2}) and from the fact that the polarized and 
unpolarized parton distributions are different.  However, the contribution to the cross sections 
from ${\hat\sigma}_4$ are small, as can be seen in Fig.~8 and Fig.~9,
which show the individual K-factors arising from the four terms in (\ref{rdf_conave2}) for 
the unpolarized cross section and from the four terms in (\ref{rdf_conll2}) for the
polarized cross section, respectively.  For both the polarized and unpolarized case the 
dominant term in the K-factor is ${\tilde\sigma}_1$.

\subsection{Spin Cross Sections}

The spin cross sections $\sigma^{NS}_{++}$ and $\sigma^{NS}_{+-}$ are calculated from $\sigma_{NS}^{\Sigma}$ and
$\sigma_{NS}^{LL}$ as follows:
\beqa
\sigma^{NS}_{++}=&\sigma_{NS}^{\Sigma}+\sigma_{NS}^{LL}, \nonumber \\
\sigma^{NS}_{+-}=&\sigma_{NS}^{\Sigma}-\sigma_{NS}^{LL},
\eeqa
and are shown in Fig.~10.  To leading order at the parton level,
\beq
{\hat\sigma}^{NS}_{++}=0\quad ({\rm leading\ order}),
\label{rdf_bornpp}
\eeq
which from (\ref{rdf_new2a}) - at the hadron level - implies an ``indirect" contribution of the form
\beq
\sigma_{++}^{NS}=2\sum_{u,d}\left(G_{\uparrow\to\uparrow}^{NS}
G_{\uparrow\to\downarrow}^{NS}
+G_{\uparrow\to\downarrow}^{NS}
G_{\uparrow\to\uparrow}^{NS}\right)
\otimes {\hat\sigma}^{\Sigma}_{NS}\quad ({\rm leading\ order}),
\eeq
where
\beq
{\hat\sigma}^{\Sigma}_{NS}={\hat\sigma}_B\quad ({\rm leading\ order}).
\eeq
Fig.~11 shows that the probability of finding a quark carrying helicity opposite to that of the
proton ($G_{\uparrow\to\downarrow}$ or $G_{\downarrow\to\uparrow}$) is not small especially 
at small $x$ and hence at the hadron level $\sigma_{++}^{NS}$ is not zero.  

At order $\alpha_s^2$,
\beqa
\sigma_{++}^{NS}=&-2\sum_{u,d}\left(G_{\uparrow\to \uparrow}^{NS}
G_{\uparrow\to\uparrow}^{NS}
+G_{\uparrow\to\downarrow}^{NS}
G_{\uparrow\to\downarrow}^{NS}\right)
\otimes {\hat\sigma}_{4} \nonumber \\
&+2\sum_{u,d}\left(G_{\uparrow\to\uparrow}^{NS}
G_{\uparrow\to\downarrow}^{NS}
+G_{\uparrow\to\downarrow}^{NS}
G_{\uparrow\to\uparrow}^{NS}\right)
\otimes( {\hat\sigma}^{\Sigma}_{NS}+{\hat\sigma}_{4}),
\eeqa
where ${\hat\sigma}^{\Sigma}_{NS}$ is given in (\ref{rdf_conave2}).  Now there is a ``direct"
contribution to $\sigma_{++}^{NS}$ coming from ${\hat\sigma}_{4}$.  However, this contribution is
small and the primary higher order contribution comes from the ${\tilde\sigma}_{1}$ correction to
${\hat\sigma}^{\Sigma}_{NS}$ in the ``indirect" contribution.

\subsection{Spin Asymmetry $A_{LL}$}

The non-singlet spin asymmetry parameter $A^{NS}_{LL}$ is defined by,
\beq
A^{NS}_{LL}={\sigma_{++}^{NS}-\sigma_{+-}^{NS}\over \sigma_{++}^{NS}+\sigma_{+-}^{NS}}
={\sigma_{NS}^{LL}\over \sigma_{NS}^{\Sigma}},
\eeq
and from (\ref{rdf_conave1}) and (\ref{rdf_conll1}) see that
\beq
A^{NS}_{LL}={{\sum_{q=u,d}\left(\Delta G_{p\to q}^{NS}\Delta G_{p\to q}^{NS}\right)\otimes {\hat\sigma}_{NS}^{LL}}
\over
{\sum_{q=u,d}\left(G_{p\to q}^{NS}G_{p\to q}^{NS}\right)\otimes {\hat\sigma}_{NS}^{\Sigma}} }.
\label{rdf_conall}
\eeq
Equation (\ref{rdf_bornpp}) implies that to leading order at the parton level,
\beq
{\hat\sigma}_{NS}^{LL}=-{\hat\sigma}_{NS}^{\Sigma}
\ {\rm and}\ 
{\hat A}^{NS}_{LL}=-1\quad ({\rm leading\ order}).
\label{rdf_losig}
\eeq
We can make a ``zero order" estimate of $A^{NS}_{LL}$ by simply ignoring the convolution
in (\ref{rdf_conall}) and using (\ref{rdf_losig}) which results in
\beq
A^{NS}_{LL}({\rm order}\ 0)=-{(\nicefrac49)\Delta G_{p\to u}^{NS}\Delta G_{p\to u}^{NS}+(\nicefrac19)\Delta G_{p\to d}^{NS}\Delta G_{p\to d}^{NS}
\over
(\nicefrac49)G_{p\to u}^{NS}G_{p\to u}^{NS}+(\nicefrac19)G_{p\to d}^{NS}G_{p\to d}^{NS}}.
\label{rdf_order0}
\eeq
Fig.~12 shows this ratio of parton distributions at $Q=M=10\gev$ versus $x$. 
The result is compared with the complete
leading order and beyond leading order predictions for $A^{NS}_{LL}$ from (\ref{rdf_conall}), integrated over $y$ 
at $W=200\gev$ plotted versus $x_T$.  At $Q=10\gev$ the leading order and the beyond leading order
results are almost identical and both resemble the zero order estimate.

Fig.~13 shows the leading and beyond leading order results for the non-singlet asymmetry parameter
$A^{NS}_{LL}$ for proton-proton collisions at $W=200\gev$ and $Q=M=2\gev$.  The asymmetry is around
$15\%$ between $p_T$ of $2$ and $8\gev$ and there is very little difference between the leading
and beyond leading order.  As can be seen in Fig.~6 and Fig.~7, the K-factor for the asymmetry parameter
$A^{NS}_{LL}$ is much smaller than the K-factors for unpolarized or polarized cross sections,
$\sigma_{\Sigma}^{LL}$ and $\sigma_{NS}^{LL}$.

\section{Summary and Conclusions}

{}From (\ref{rdf_conave2}) and (\ref{rdf_conll2}) we see that to order $\alpha_s^2$ at the
parton level,
\beq
-{\hat\sigma}_{NS}^{LL}={\hat\sigma}_{NS}^{\Sigma}+2{\hat\sigma}_{4},
\eeq
which means that except small differences due to the term ${\hat\sigma}_{4}$ and small differences arising
from the fact that the polarized parton distributions $\Delta G^{NS}_{p\to q}$ in (\ref{rdf_conll1}) differ from the unpolarized one $G^{NS}_{p\to q}$ in (\ref{rdf_conave1}), the higher order corrections to
polarized cross section, $-{\hat\sigma}_{NS}^{LL}$, and the unpolarized cross section
${\hat\sigma}_{NS}^{\Sigma}$ are identical.  In both cases the higher order corrections are large
and important.  Everything that has previously been said about the higher order corrections to the
unpolarized cross section, for example \cite{EMP}, is true for the polarized case as well.

The higher order corrections to the ratio of the polarized to the unpolarized cross section,
\beq
A^{NS}_{LL}=\sigma_{NS}^{LL}/\sigma_{NS}^{\Sigma},
\eeq
are small since the big corrections to the numerator are canceled by big corrections to the 
denominator.  As can be seen in Fig.~12 and Fig.~13, $A^{NS}_{LL}$ is very stable against 
higher order corrections.  The asymmetry is around
$15\%$ between $p_T$ of $2$ and $8\gev$ in proton-proton collisions at $W=200\gev$ and 
$Q=M=2\gev$.  The non-singlet asymmetry, $A^{NS}_{LL}$, is a direct measurement of the 
non-singlet (\ie valence) polarized quark distributions within the proton.  However, to
construct the non-singlet observables in this paper one must subtract $pp$ spin measurements
from the corresponding $\bar{p}p$ measurement, and it is unlikely that this will be done
in the near future.  On the other hand, if $A_{LL}$ is measured in $pp$ collisions at
RHIC, then deviations from our predictions will indicate large contributions from
gluons, $\Delta G_{p\to g}$, and/or from the sea, $\Delta G_{p\to s}$.  This would provide 
the first hint as to the behavior of the polarized gluon parton distribution, $\Delta G_{p\to g}$.

\centerline{\bf Acknowledgements}
We warmly thank L. E. Gordon and Cetin Savkli for illuminating discussions.

\newpage

% Figure 1
\begin{figure}
\centerline{\epsfbox{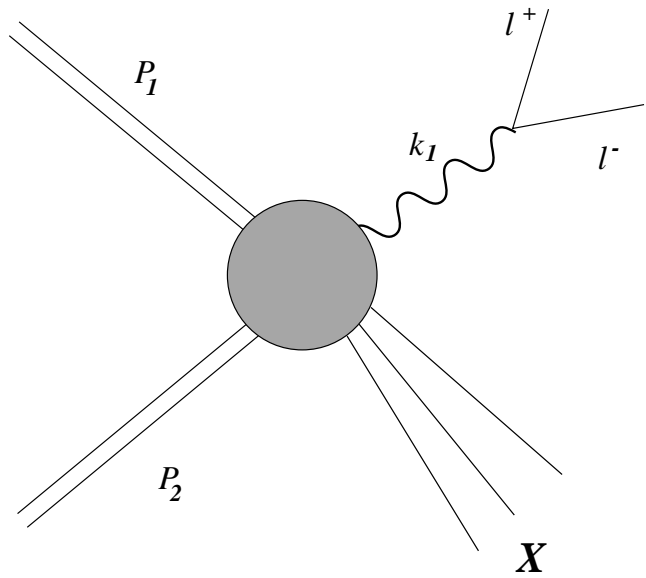}}
\caption{The Drell-Yan Process.}
\end{figure}

% Figure 2
\begin{figure}
\centerline{\epsfbox{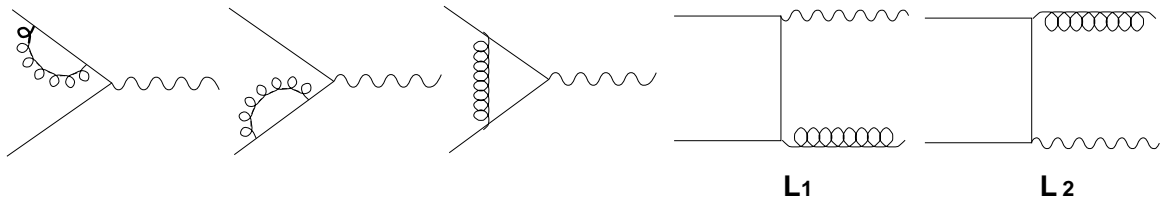}}
\caption{Radiative corrections at $p_T=0$ and the lowest order 
contributions at $p_T\neq 0$.}
\end{figure}

% Figure 3
\begin{figure}
\epsfxsize=11cm
\centerline{\epsfbox{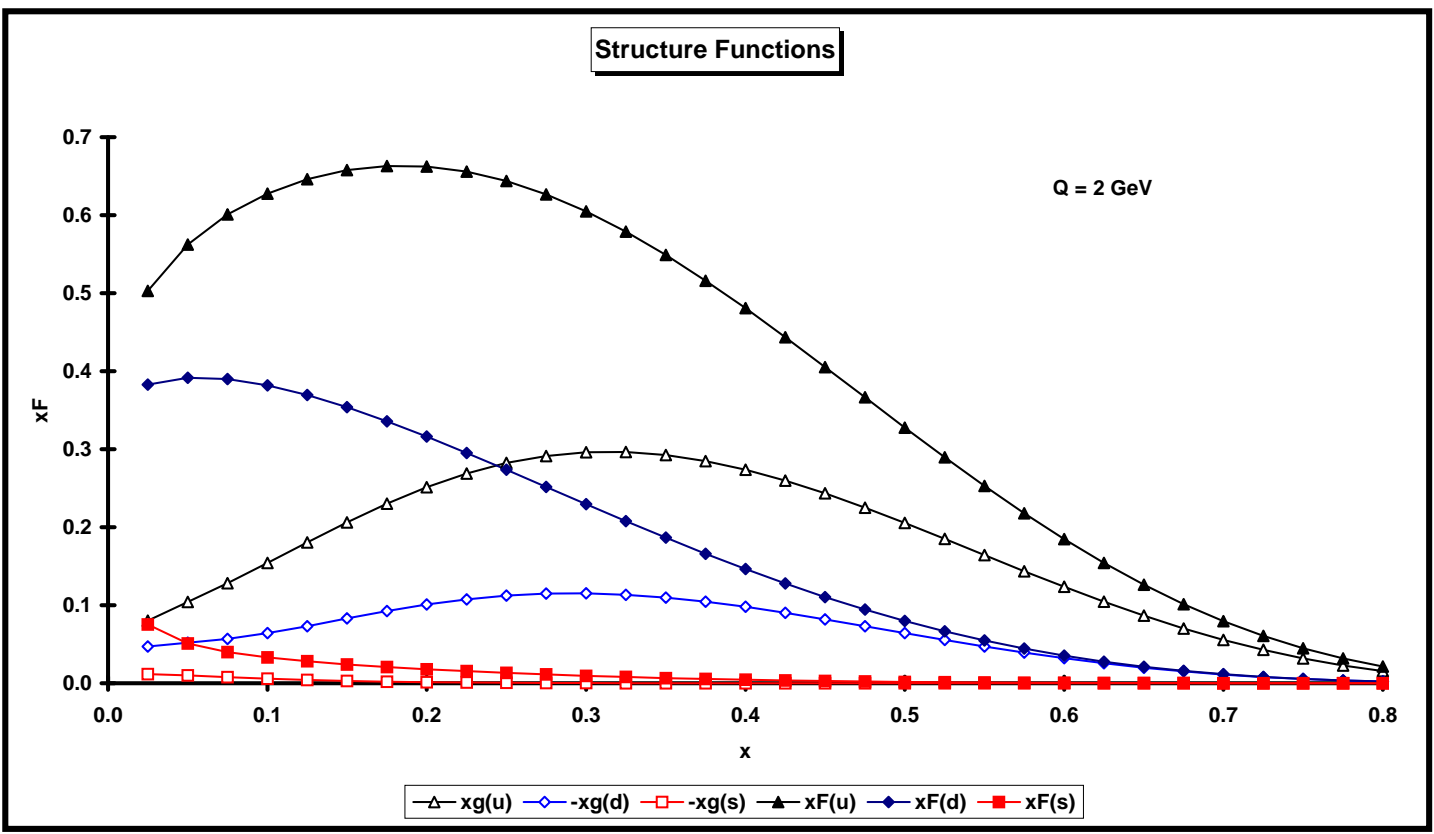}}
\caption{Polarized ($xg$) and unpolarized ($xF$) structure functions
at $Q=2$ GeV from Ref.\cite{GS} (LO solution A).}
\end{figure}

% Figure 4
\begin{figure}
\epsfxsize=14cm
\centerline{\epsfbox{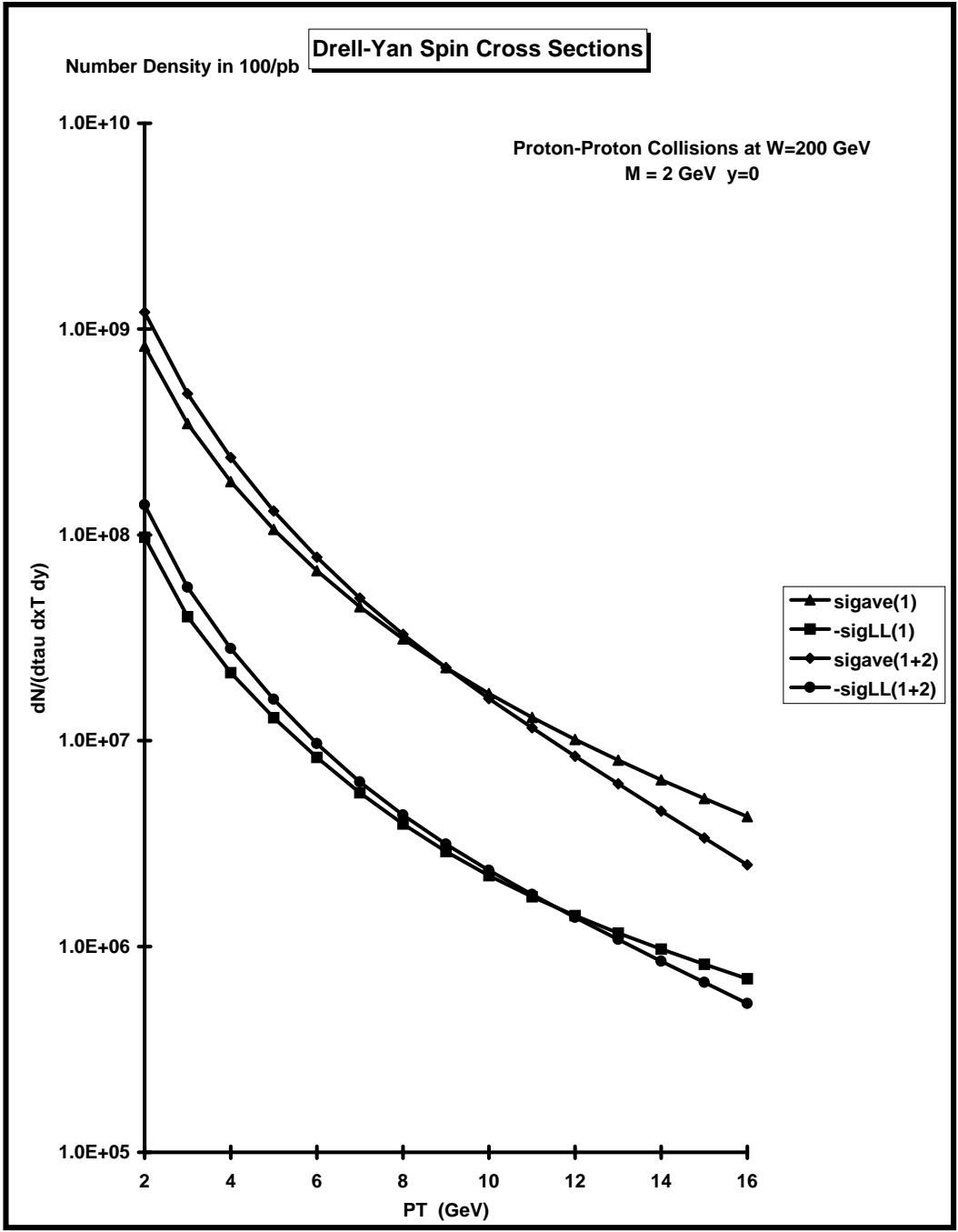}}
\caption{The non-singlet unpolarized spin-averaged (ave) cross section, $\sigma_{NS}^{\Sigma}$,
and longitudinally polarized (LL) cross section, $\sigma_{NS}^{LL}$, for $pp$ 
collisions at $W=200\gev$,
$Q=M=2\gev$, and $y=0$ as a function of $p_T$.  The plot shows the ``number density", 
$dN/d\tau dx_Tdy$, in $100/$pb at leading order (1) and at order $\alpha_s^2$ (1+2).}
\end{figure}

% Figure 5
\begin{figure}
\epsfxsize=14cm
\centerline{\epsfbox{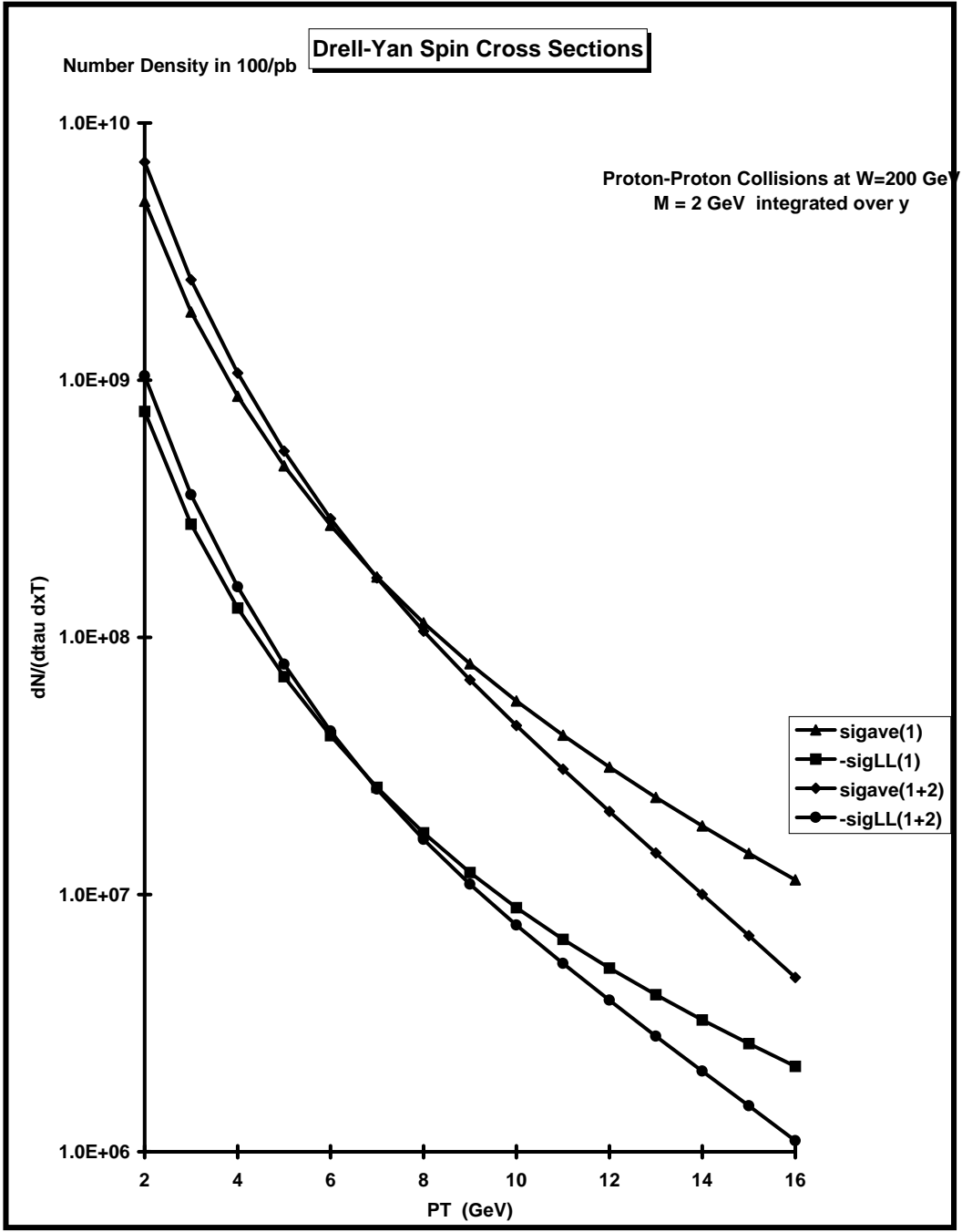}}
\caption{
Same as Fig.~4 integrated over rapidity $y$.}
\end{figure}

% Figure 6
\begin{figure}
\epsfxsize=10cm
\centerline{\epsfbox{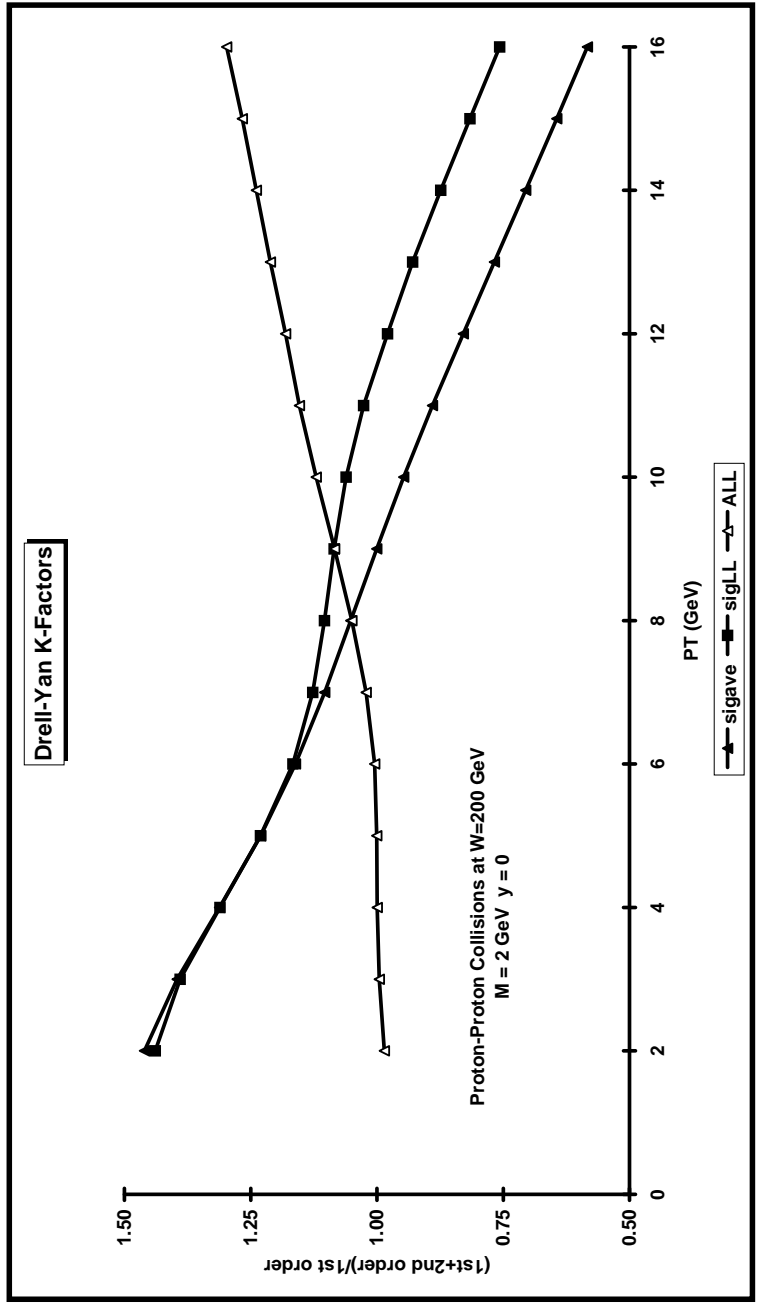}}
\caption{
Shows the $K$-factors for non-singlet unpolarized spin-averaged (ave) 
and longitudinally polarized (LL) cross sections for $pp$ collisions at $W=200\gev$,
$Q=M=2\gev$, and $y=0$ as a function of $p_T$ in Fig.~4.  The $K$-factor is the 
ratio of the full order $\alpha_s^2$ cross section (1+2) to the leading order
Born term (1). Also shown is the $K$-factor for the non-singlet spin symmetry 
parameter $A_{LL}^{NS}=\sigma_{NS}^{LL}/\sigma_{NS}^{\Sigma}$.  }
\end{figure}

% Figure 7
\begin{figure}
\epsfxsize=10cm
\centerline{\epsfbox{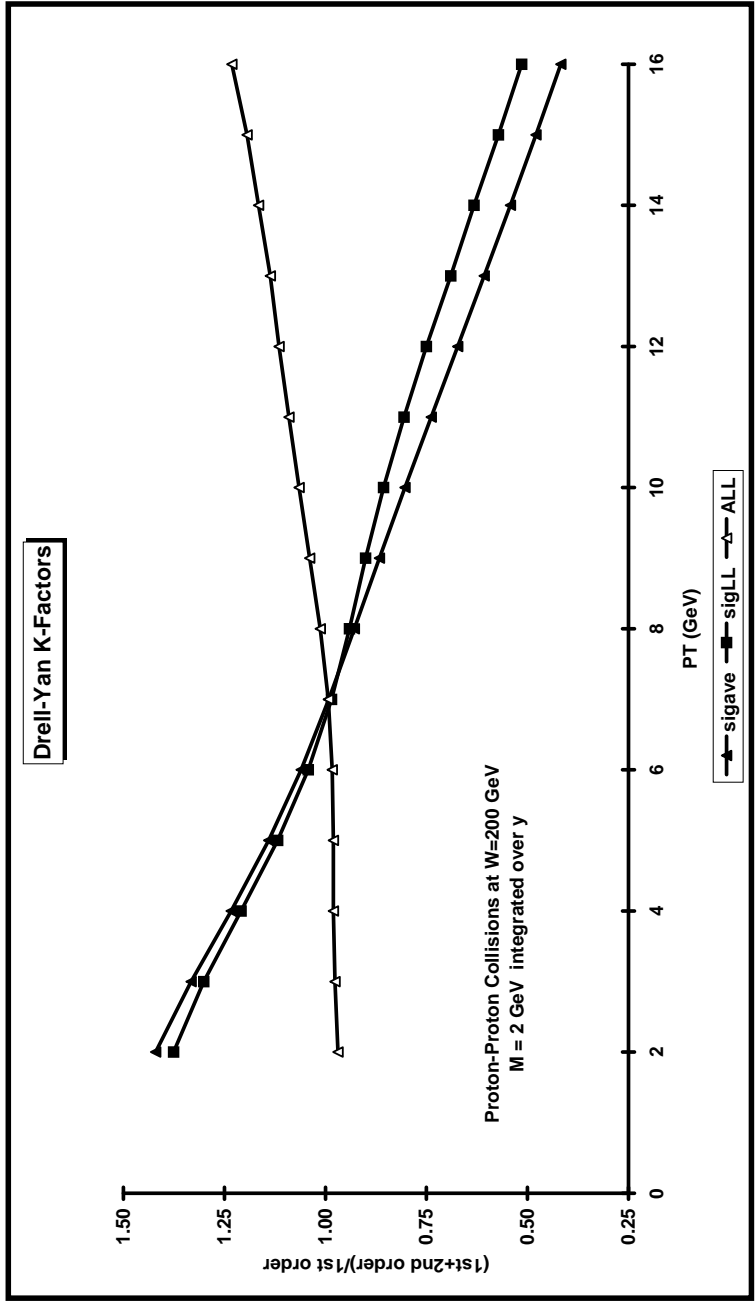}}
\caption{
Shows the $K$-factors for non-singlet unpolarized spin-averaged (ave) 
and longitudinally polarized (LL) cross sections for $pp$ collisions at $W=200\gev$,
$Q=M=2\gev$, integrated over rapidity $y$ as a function of $p_T$ in Fig.~5.  The $K$-factor is the 
ratio of the full order $\alpha_s^2$ cross section (1+2) to the leading order
Born term (1). Also shown is the $K$-factor for the non-singlet spin symmetry 
parameter $A_{LL}^{NS}=\sigma_{NS}^{LL}/\sigma_{NS}^{\Sigma}$.  }
\end{figure}

% Figure 8
\begin{figure}
\epsfxsize=10cm
\centerline{\epsfbox{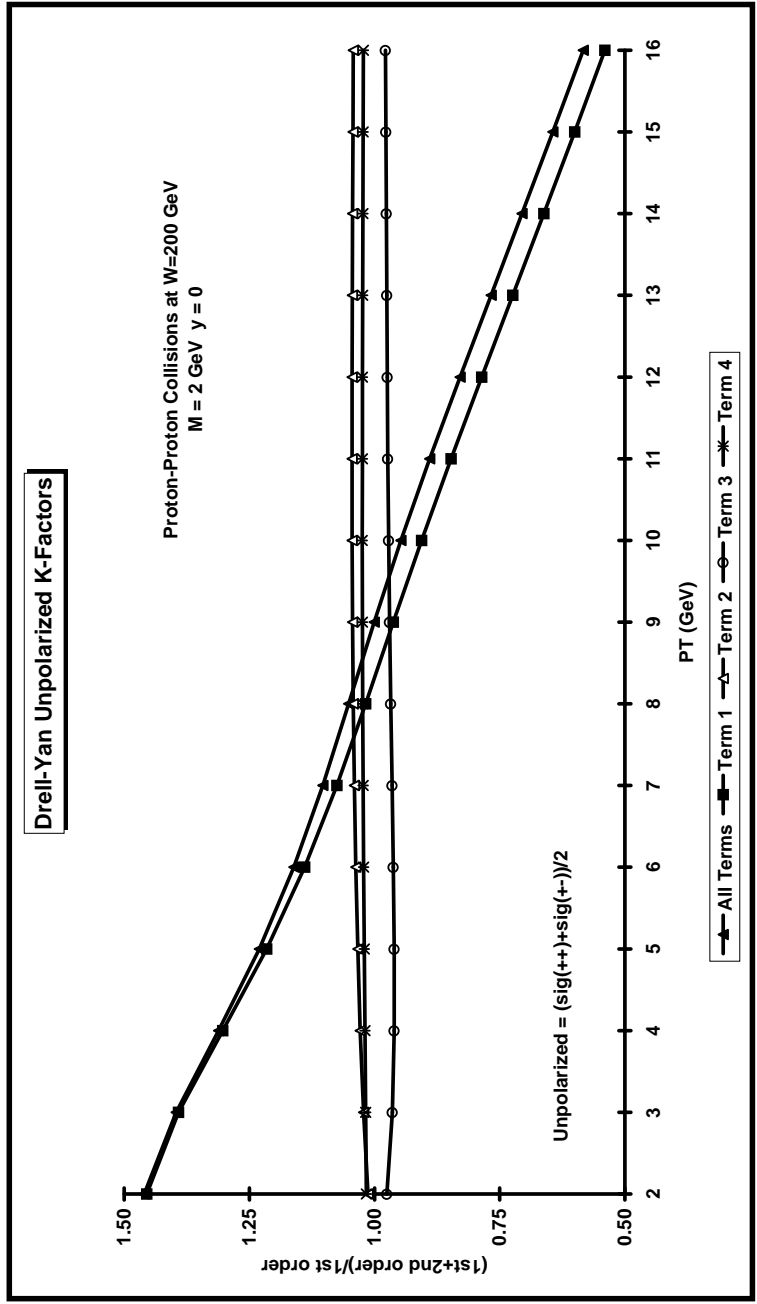}}
\caption{
Shows the $K$-factors arising from each of the four terms in (\ref{rdf_conave2})
for non-singlet unpolarized spin-averaged cross section, $\sigma_{NS}^{\Sigma}$, 
for $pp$ collisions at $W=200\gev$,
$Q=M=2\gev$, and $y=0$ as a function of $p_T$.  The $K$-factor is the 
ratio of the order $\alpha_s^2$ cross section to the leading order
Born term. }
\end{figure}

% Figure 9
\begin{figure}
\epsfxsize=10cm
\centerline{\epsfbox{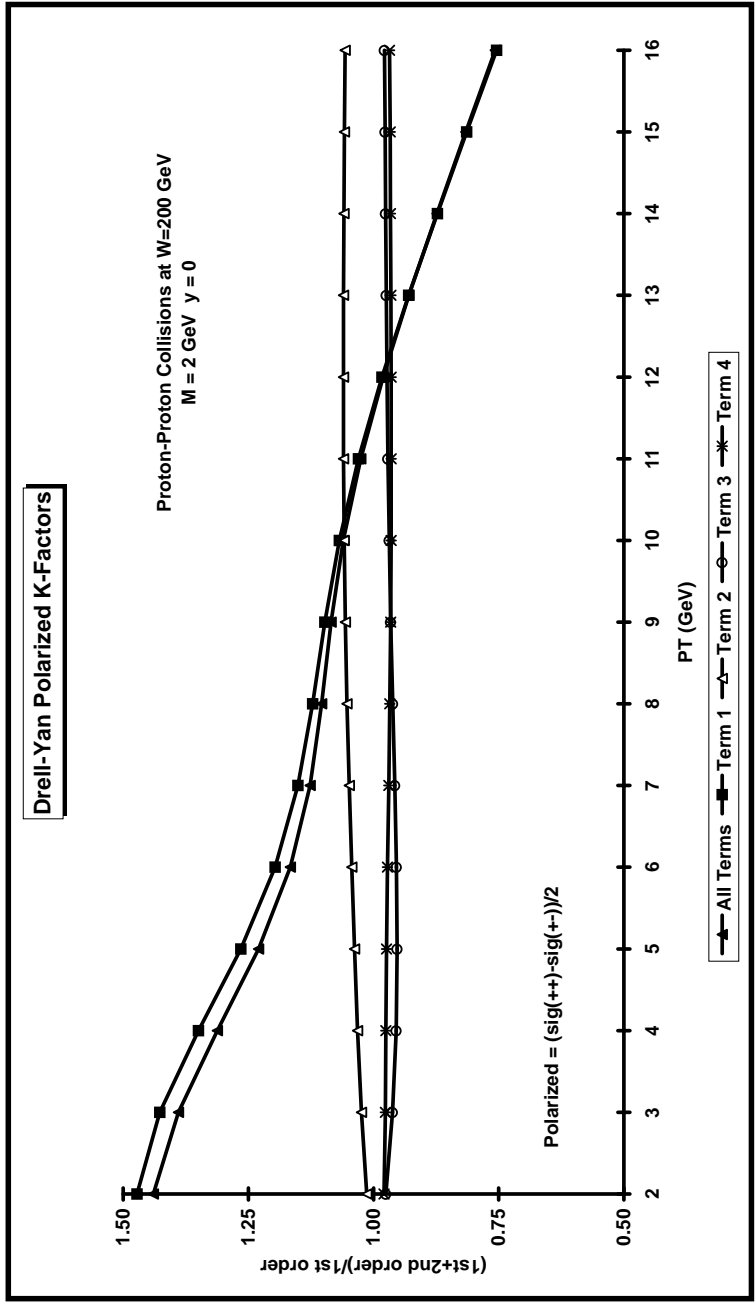}}
\caption{
Shows the $K$-factors arising from each of the four terms in (\ref{rdf_conll2})
for non-singlet longitudinally polarized (LL) cross section, $\sigma_{NS}^{LL}$,
for $pp$ collisions at $W=200\gev$,
$Q=M=2\gev$, and $y=0$ as a function of $p_T$.  The $K$-factor is the 
ratio of the order $\alpha_s^2$ cross section to the leading order Born term. }
\end{figure}

% Figure 10
\begin{figure}
\epsfxsize=14cm
\centerline{\epsfbox{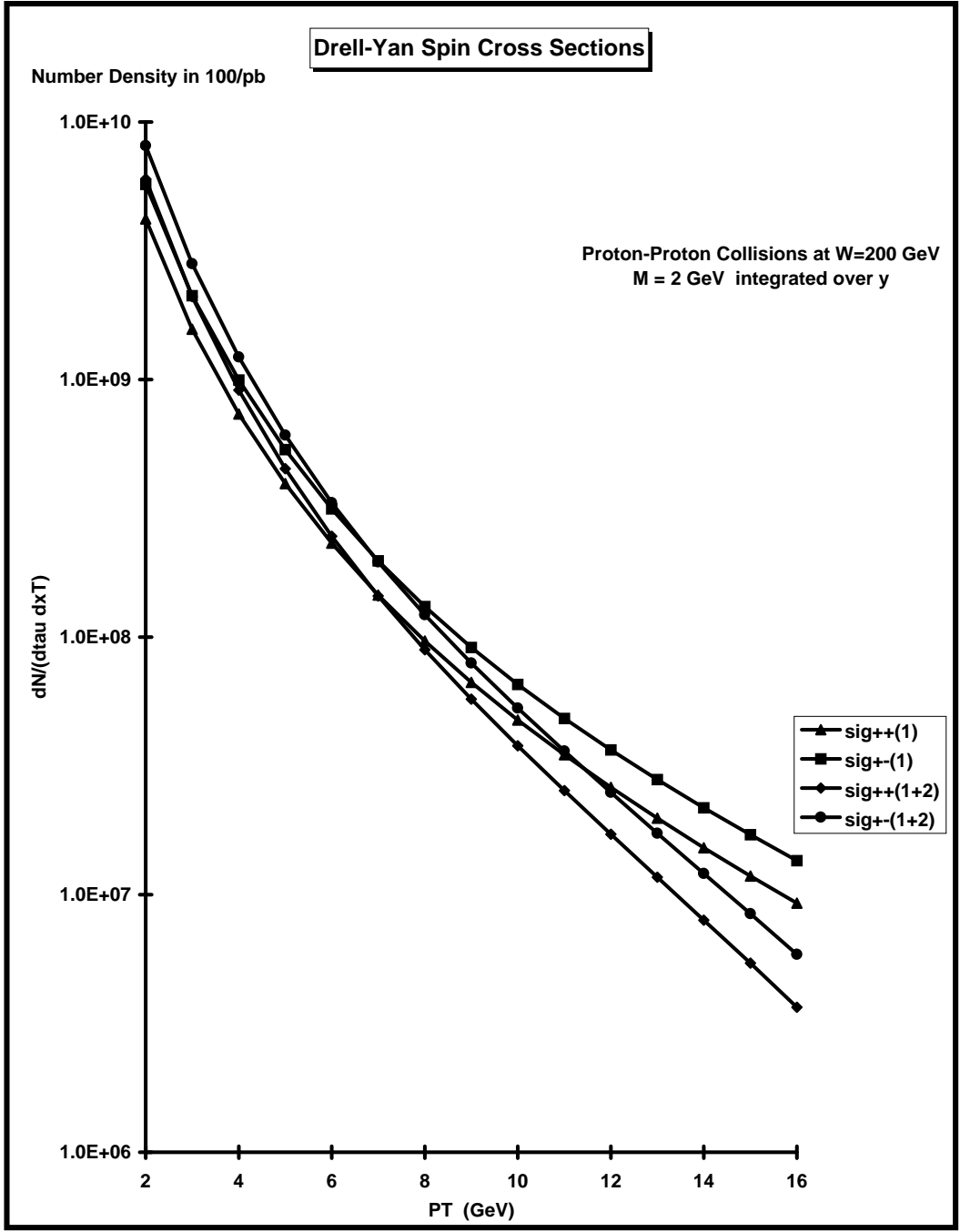}}
\caption{
The non-singlet spin cross sections, $\sigma^{NS}_{++}$ and $\sigma^{NS}_{+-}$,
for $pp$ collisions at $W=200\gev$,
$Q=M=2\gev$, integrated over $y$ as a function of $p_T$.  The plot shows the ``number density", 
$dN/d\tau dx_T$, in $100/$pb at leading order (1) and at order $\alpha_s^2$ (1+2).}
\end{figure}

% Figure 11
\begin{figure}
\epsfxsize=10cm
\centerline{\epsfbox{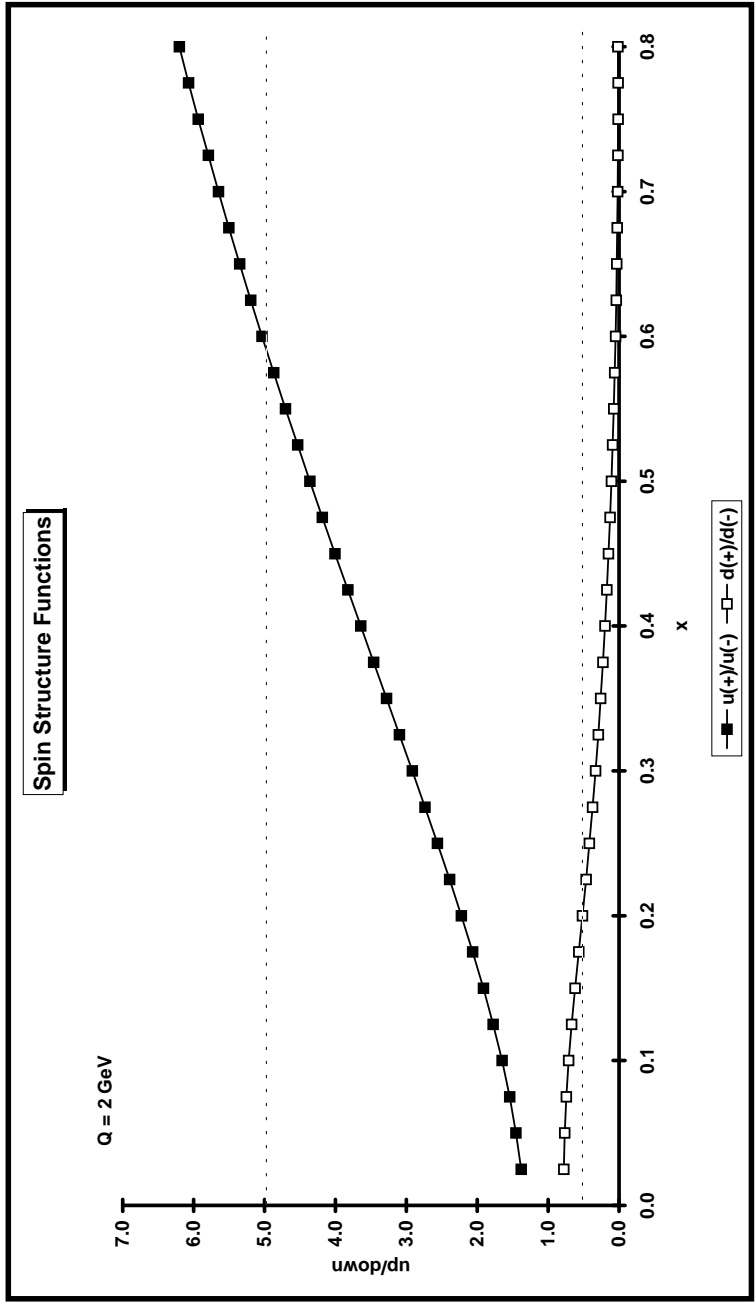}}
\caption{
Shows the ratio, $G_{\uparrow\to\uparrow}$/$G_{\uparrow\to\downarrow}$, for $u$ and
$d$ quarks at $Q=2\gev$ versus $x$ resulting from the polarized structure functions in Fig.~3.}
\end{figure}

% Figure 12
\begin{figure}
\epsfxsize=10cm
\centerline{\epsfbox{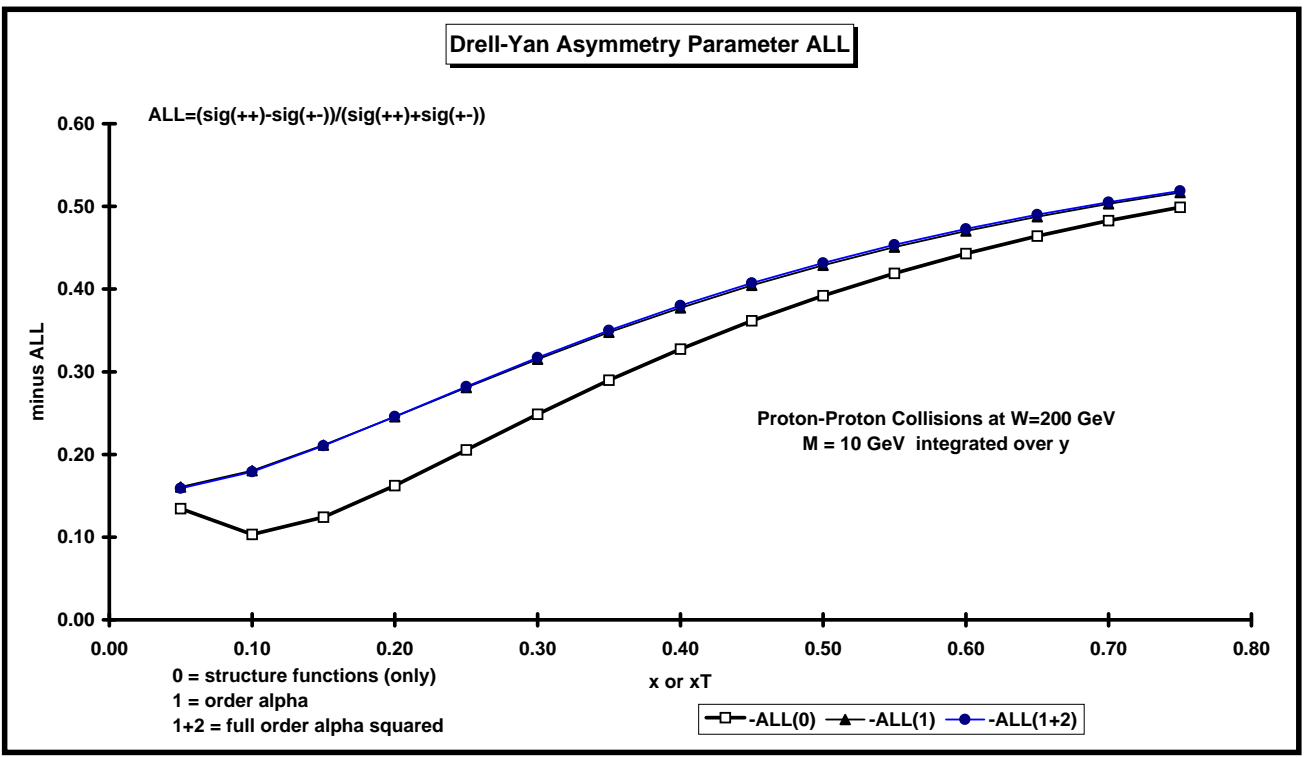}}
\caption{
The Drell-Yan non-singlet spin asymmetry parameter 
$-A_{LL}^{NS}=-\sigma_{NS}^{LL}/\sigma_{NS}^{\Sigma}$ for $pp$ collisions at $W=200\gev$,
$Q=M=10\gev$, integrated over $y$ as a function of $x_T$ at at leading order (1) 
and at order $\alpha_s^2$ (1+2).  Also shown is the ``zero order" structure function 
estimate in (\ref{rdf_order0}) plotted versus $x$ (0). }
\end{figure}

% Figure 13
\begin{figure}
\epsfxsize=10cm
\centerline{\epsfbox{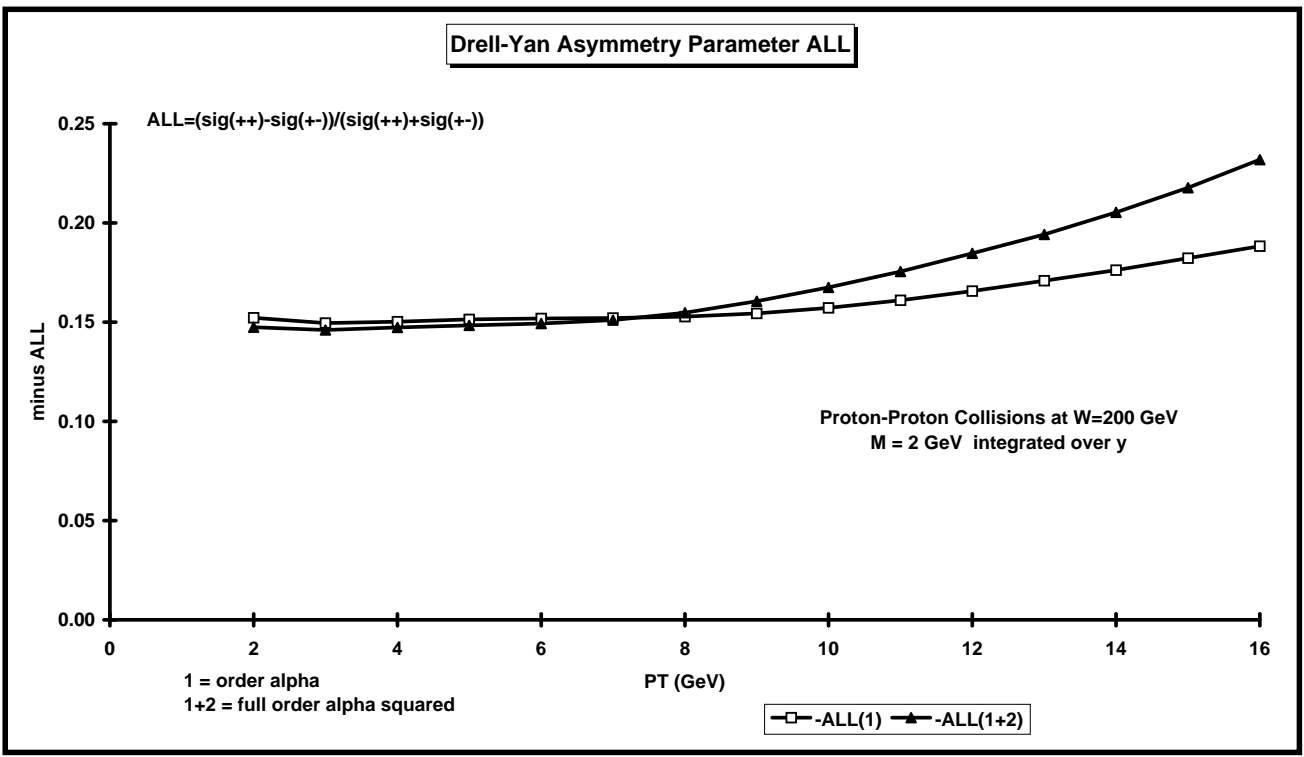}}
\caption{
The Drell-Yan non-singlet spin asymmetry parameter 
$-A_{LL}^{NS}=-\sigma_{NS}^{LL}/\sigma_{NS}^{\Sigma}$ for $pp$ collisions at $W=200\gev$,
$Q=M=2\gev$, integrated over $y$ as a function of $p_T$ at at leading order (1) 
and at order $\alpha_s^2$ (1+2).  }
\end{figure}

\end{document}